\documentclass[a4paper,11pt]{article}
\usepackage{pos}
\usepackage{comment}
\title{Global polarization and spin alignment in heavy-ion collisions: past, present and future}

\author[a]{Wen-Bo Dong}
\author[b]{Xin-Li Sheng}
\author[a]{Yi-Liang Yin}
\author*[a,c]{Qun Wang}

\affiliation[a]{Department of Modern Physics, University of Science and Technology of China,\\ 
Hefei, Anhui 230026, China}

\affiliation[b]{
INFN Sezione di Firenze, Via Giovanni Sansone 1, 50019, Sesto Fiorentino FI, Italy}

\affiliation[c]{
School of Mechanics and Physics, Anhui University of Science and Technology, \\
Huainan,Anhui 232001, China}

\emailAdd{wenba@mail.ustc.edu.cn}
\emailAdd{sheng@fi.infn.it}
\emailAdd{yinyiliang@mail.ustc.edu.cn}
\emailAdd{qunwang@ustc.edu.cn}

\abstract{We give a brief overview on global polarization and spin alignment in heavy ion collisions. The current theoretical understandings on global polarization of hyperons and the global spin alignment of vector mesons are summarized.}

\FullConference{25th International Spin Physics Symposium (SPIN 2023)\\
 24-29 September 2023\\
 Durham, NC, USA\\}

\begin{document}
\maketitle

\section{Introduction}
Non-central heavy-ion collisions induce a huge global orbital angular momentum (OAM) perpendicular to the reaction plane. Due to the spin-orbit coupling, the initial OAM is partially converted to the spin angular momentum of particles in the quark-gluon plasma, in a similar way as the Barnett effect \cite{barnett1915magnetization} and the Einstein-de Hass effect \cite{einstein1915experimental}. This in turn results in phenomena such as polarizations of $\Lambda$ hyperons \cite{STAR:2017ckg,STAR:2018gyt,STAR:2019erd} and vector mesons \cite{ALICE:2019aid,ALICE:2020iev,STAR:2022fan,ALICE:2022dyy}, which have been measured in recent experiments. In this proceeding, we reviewed recent progresses on the polarization phenomena in heavy-ion collisions and discussed our present understanding about the vector meson's polarization. We also proposed questions for possible future studies. One can also read \cite{Wang:2017jpl,Becattini:2020ngo,Huang:2020dtn,Gao:2020lxh,Becattini:2022zvf,Hidaka:2022dmn} for recent reviews on polarizations in heavy-ion collisions.

\section{Global polarization}\label{sec:polarization}

The idea was first proposed by Z.-T. Liang and X.-N. Wang in 2004 that the huge OAM in non-central heavy-ion collisions can polarize quarks due to the spin-orbit coupling \cite{Liang:2004xn,Liang:2004ph}. They studied the parton scatterings within the screened potential model. Under the small angle approximation, the cross section can be decomposed into a spin-dependent part and a spin-independent part, while the former part is proportional to the spin-orbit coupling ${\bf n}\cdot({\bf x}_T\times{\bf p})$, where $\bf n$ is the spin quantization direction, ${\bf x}_T$ is the impact parameter relative to the center of the static potential, and ${\bf p}$  is the momentum of incoming parton. Such a coupling indicates an unpolarized quark will be polarized along the direction of local OAM, i.e., the direction of ${\bf x}_T\times{\bf p}$, after scattering with the static potential. In non-central heavy-ion collisions, the initial geometry leads to the fact that the average longitudinal momentum per parton depends on the position in $x$-direction, described by a nonzero $dp_z/dx$. Microscopically, the local OAM in parton collisions, given by $dp_z dx$, has a preferred direction along $-y$, which is the same as the direction of the global OAM. Numerical similations based on Woods-Saxon model and hard-sphere model \cite{Liang:2004ph,Gao:2007bc} show that the global OAM in Au+Au collisions at 200 GeV can reach $10^5\hbar$, while the local OAM can be as large as $1\sim 4\,\hbar$, leading to the global quark polarization perpendicular to the reaction plane. 

Instead of treating one quark as an effective potential, the polarized quark scatterings are studied in \cite{Gao:2007bc} in a more realistic way by treating two quarks as plane waves and consider the scattering through one-gluon exchange. In order to recover the impact parameter, they made a two dimensional Fourier transformation with respect to the transverse momentum transfer. The result shows that the spin-dependent part of the cross section is proportional to the spin-orbit coupling, which qualitatively agrees with the result obtained by the potential model \cite{Liang:2004ph}. The model was further improved in Ref. \cite{Zhang:2019xya} by parameterizing the incident particles as wave packets with certain momenta and center positions. The initial nonvanishing OAM is included by a finite transverse distance (impact parameter) between centers of two wave packets. For simplicity, the outgoing particles are treated as plane waves with zero OAM. Therefore the initial OAM is fully converted to the spin polarization of outgoing particles. The authors in \cite{Zhang:2019xya} also imposed the causality condition in the center of mass frame for each parton collision, which declared that the collision happens at the same time, but the incident particles are displaced by the impact parameter. The cross section contains a 16-dimensional integral, which is challenging for numerical calculations. The authors developed a Monte-Carlo integration package running on multi-GPUs \cite{Wu:2019tsf}, which makes the integral possible within acceptable computing time. Another challenge in the model is that the polarized squared amplitude contains more than 5000 terms, making it hard to see the physics behind the formula. Fortunately, the polarization production rate has a concise form as $d^4{\bf P}_q/d^4x\propto \boldsymbol{\nabla}\times (\beta {\bf u})$ when the partons' momentum is in local equilibrium while their spin is not. Here $\beta$  is the inverse of temperature and $\bf u$ is the local fluid velocity. This work \cite{Zhang:2019xya} therefore provides a microscopic description for the generation of global spin polarization, which is based on a comprehensive first principle calculation including all possible parton scattering channels. 

The kinetics of spin 1/2 particles can be analytically described by the Wigner function approach \cite{Elze:1986qd,Vasak:1987um,Hidaka:2016yjf,Weickgenannt:2019dks,Hattori:2019ahi,Gao:2019znl,Wang:2019moi,Weickgenannt:2020aaf,Yang:2020hri,Sheng:2021kfc}, see Ref. \cite{Gao:2020vbh,Huang:2020dtn,Hidaka:2022dmn} for recent reviews. The axial vector component of the Wigner function denotes the quark polarization density. In general, the polarization at local equilibrium contains various sources at linear order in gradient, including the thermal vorticity, thermal shear, fluid acceleration, and electromagnetic fields \cite{Hidaka:2017auj,Yi:2021ryh}. Among these sources, the thermal vorticity is dominant for $\Lambda$'s global polarization, while the thermal shear plays the most important role for the $\Lambda$'s longitudinal polarization \cite{Becattini:2021suc,Becattini:2021iol,Liu:2021uhn,Fu:2021pok}. However, how the quark-gluon plasma reaches local equilibrium with spin still remains a puzzle. For massive fermions, the kinetic theory becomes messy because of increasing degrees of freedom (DOF). Scalar particles are described by one Boltzmann equation, while chiral fermions are described by the chiral kinetic theory, containing two equations for left and right-handed fermions, respectively.  However, the kinetics for massive fermions are described by the Kadanoff-Baym equation \cite{kadanoff2018quantum,Mrowczynski:1992hq} for the Wigner function, which contains 16 DOF.  Fortunately, the semi-classical expansion approach provides a systematic way to derive the Wigner function order by order, which reduces the independent DOF to 4 at each order. In this case, particles are characterized by the matrix-valued spin dependent distributions (MVSD),
\begin{equation}
f^q_{rs}(x,p)\equiv\int\frac{d^4q}{2(2\pi)^3} e^{-iq\cdot x}\delta(p\cdot q)\left\langle a^\dagger(s,p_2)a(r,p_1)\right\rangle\,,
\end{equation}
where $a^\dagger,a$ are creation and annihilation operators, respectively, and $p_{1,2}\equiv p\pm q/2$. The trace in spin space of the MVSD is interpreted as the unpolarized distributions, while its projection onto Pauli matrices are related to the spin polarization as 
\begin{equation}
f^q_{rs}(x,p)=\frac{1}{2}f_q(x,p)\left[\delta_{rs}-P^q_{\mu}(x,p)n_{j}^{(+)\mu}(x,p)\tau_{rs}^{j}\right]\,, \label{eq:quark dis}
\end{equation}
where $P^{q}_\mu$  is the polarization vector, $n_{j}^{(+)\mu}$ is the Lorentz transformed four-vector of the basis direction ${\bf n}_j$ in the rest frame of particle, and $\tau^j$  is the Pauli matrix corresponding to the direction ${\bf n}_j$. In terms of the MVSD, the spin Boltzmann equations (SBEs) can be derived from the Kadanoff-Baym equation. At zeroth order in $\hbar$, the SBEs read \cite{Sheng:2021kfc}
\begin{eqnarray}
\frac{1}{E_{p}}p\cdot\partial_{x}\text{tr}\left[f^{(0)}(x,p)\right] & = & \mathscr{C}_\text{scalar}[f^{(0)}]\,,\nonumber \\
\frac{1}{E_{p}}p\cdot\partial_{x}\text{tr}\left[n_{j}^{(+)\mu}\tau_{j}f^{(0)}(x,p)\right] & = & \mathscr{C}_\text{pol}[f^{(0)}]\,, \label{eq:LO SBE}
\end{eqnarray}
in which only leading order MVSD $f^{(0)}$ appears. Explicit expressions for collision terms $\mathscr{C}_\text{scalar}$ and $\mathscr{C}_\text{pol}$ can be found in Ref. \cite{Sheng:2023urn}. At first order in $\hbar$, the SBEs are written as \cite{Sheng:2023urn}
\begin{eqnarray}
\frac{1}{E_{p}}p\cdot\partial_{x}\text{tr}\left[f^{(1)}(x,p)\right] & = & \mathscr{C}_\text{scalar}[f^{(1)},\partial f^{(0)},f^{(0)}]\,,\nonumber \\
\frac{1}{E_{p}}p\cdot\partial_{x}\text{tr}\left[n_{j}^{(+)\mu}\tau_{j}f^{(1)}(x,p)\right] & = & \mathscr{C}_\text{pol}[f^{(1)},\partial f^{(0)},f^{(0)}]\,.\label{eq:NLO SBE}
\end{eqnarray}
The collision terms on the right-hand side in general depend not only on the MVSD, but also on the space-time derivate of MVSD, $\partial^\mu f^{(0)}$. Microscopically, this reflects the spin-orbit coupling during particle scatterings, which is the source of spin polarization. Equations \ref{eq:LO SBE} and \ref{eq:NLO SBE} provide a possible way to numerically simulate the evolution of spin polarization. One can also express the kinetic equations in terms of the distribution function in extended phase space by treating spin as an additional phase space parameter \cite{Weickgenannt:2020aaf,Weickgenannt:2021cuo}, or in terms of the axial current \cite{Yang:2020hri}.

\section{Spin alignment}
\begin{figure}
\begin{center}
\includegraphics[scale=0.6]{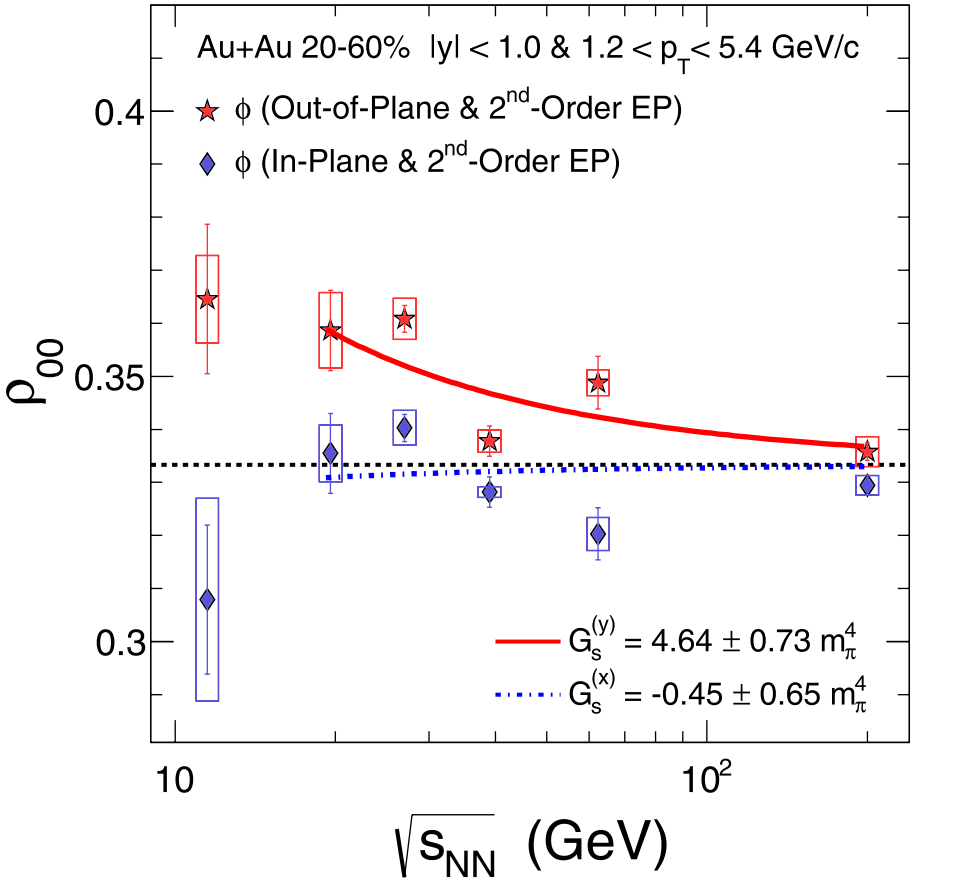}
\end{center}

\caption{The STAR's measurement on vector mesons' $\rho_{00}$ with respect to the second-order event  plane in heavy ion collision \cite{STAR:2022fan} The stars and circles  correspond to $\phi$ and $K^{*0}$  mesons, respectively. The red solid line is the fit curve using the model in Ref. \cite{Sheng:2019kmk}.} \label{fig:rho00}
\end{figure}

For vector mesons, studies usually focus on the spin alignment instead of the spin polarization. This is because vector mesons mainly decay through strong decays or dilepton decays, which preserve the parity symmetry and make it difficult to measure the spin polarization. One of the experimental observables for vector meson is the $00$-element of the spin density matrix,  i.e. $\rho_{00}$,  known as the spin alignment of vector mesons. As spin-1 particles, vector meson's spin can be $0$ and $\pm 1$, thus $\rho_{00}=1/3$ if the spin does not have any preferred direction. A positive (negative) deviation from 1/3 means the vector meson's spin vector is preferred to align parallel (perpendicular) to the direction of spin quantization direction. It was first proposed by Z.-T. Liang and X.-N. Wang in 2004 that the spin alignment is related to the polarization of constituent quark and antiquark \cite{Liang:2004xn}. Since quarks can be globally polarized by the OAM in non-central collisions, it is expected that the spin alignment measured along the direction of global OAM (which is referred as the global spin alignment) also have nonvanishing deviation from 1/3. Recently, the STAR collaboration measured the spin alignment of $\phi$ and $K^{*0}$ mesons at RHIC energies \cite{STAR:2022fan}. The result is shown in Fig. (\ref{fig:rho00}), where the spin alignment of $K^{*0}$ is consistent with 1/3, while the spin alignment of $\phi$  shows a significant positive derivation from 1/3. Experiments at LHC also observed global OAM for $\phi$ , $K^{*0}$, and $J/\psi$ in Pb-Pb collisions at $\sqrt{s_\text{NN}}=2.76$ TeV \cite{ALICE:2019aid,ALICE:2020iev,ALICE:2022dyy}.

It was recently found that the ordinary mechanism for quark spin polarization can not explain the significant spin alignment for $\phi$ meson \cite{Sheng:2019kmk}. For example, the thermal vorticity field in heavy-ion collisions is of order $10^{-2}$, therefore a quick estimation shows that the induced spin alignment, which is $\propto \omega^2$, is $\sim 10^{-4}$ and is much smaller than $\rho_{00}$ observed in experiments. A new mechanism of strong force field (effective vector field) effect is then introduced in \cite{Sheng:2019kmk,Sheng:2022wsy,Sheng:2022ffb,Sheng:2023urn} to solve this puzzle. This field refers to the strong interaction between constituent quark (antiquark) inside the meson and the quark (antiquark) in the surrounding medium. The field is dominated by fluctuations and therefore has vanishing mean value. However, it has finite contribution to the spin alignment of a flavorless vector meson, because the spin polarization for constituent quarks and antiquarks can have strong correlation. The interplay between quark and the strong force field can be effectively described by the chiral quark model, with the Lagrangian having  $SU_{L}(3)\times SU_{R}(3)$ symmetry \cite{Manohar:1983md,Fernandez:1993hx}
\begin{eqnarray}
\mathcal{L} & = & \overline{\psi}\left[i\gamma_{\mu}\left(\partial^{\mu}+igG^{\mu}\right)+g_{V}\gamma_{\mu}V^{\mu}\right]\psi+g_{A}\overline{\psi}\gamma_{\mu}A^{\mu}\psi\nonumber \\
 &  & +\frac{1}{4}f^{2}\text{Tr}\left(\partial^{\mu}\Sigma^{\dagger}\partial_{\mu}\Sigma\right)-\frac{1}{2}\text{Tr}F_{\mu\nu}F^{\mu\nu}\,,\label{eq:lag}
\end{eqnarray}
where $\psi$  is the quark field and $V^{\mu}$ is the effective vector fields induced by quark currents. The interaction Lagrangian for an $s/\overline{s}$ pair coupled with the $\phi$  field is $g_V\overline{s}\gamma^\mu sV^\phi_\mu$. The vector $\phi$ field  $V_\phi^\mu$  can polarize $s/\overline{s}$  though the magnetic or electric components of the field strength tensor $F_{\rho\sigma}^{\phi}=\partial_{\rho}V_{\sigma}^{\phi}-\partial_{\sigma}V_{\rho}^{\phi}$, in a similar way as the classical electromagnetic field. The corresponding $\phi$ meson's spin alignment in the rest frame is given by \cite{Sheng:2022wsy}
\begin{eqnarray}
\rho_{00}\left(x,\mathbf{0}\right) & \approx & \frac{1}{3}+C_{1}\left[\frac{1}{3}\omega^{\prime}\cdot\omega^{\prime}-\left(\epsilon_{0}\cdot\omega^{\prime}\right)^{2}\right]\nonumber \\
 &  & +C_{2}\left[\frac{1}{3}\varepsilon^{\prime}\cdot\varepsilon^{\prime}-\left(\epsilon_{0}\cdot\varepsilon^{\prime}\right)^{2}\right]\nonumber \\
 &  & -\frac{4g_{\phi}^{2}}{m_{\phi}^{2}T_{\text{eff}}^{2}}C_{1}\left[\frac{1}{3}\mathbf{B}_{\phi}^{\prime}\cdot\mathbf{B}_{\phi}^{\prime}-\left(\epsilon_{0}\cdot\mathbf{B}_{\phi}^{\prime}\right)^{2}\right]\nonumber \\
 &  & -\frac{4g_{\phi}^{2}}{m_{\phi}^{2}T_{\text{eff}}^{2}}C_{2}\left[\frac{1}{3}\mathbf{E}_{\phi}^{\prime}\cdot\mathbf{E}_{\phi}^{\prime}-\left(\epsilon_{0}\cdot\mathbf{E}_{\phi}^{\prime}\right)^{2}\right]\,,\label{eq:rho_00}
\end{eqnarray}
where $\epsilon_{0}$ denotes the spin quantization direction,  $\mathbf{E}_{\phi}$ and $\mathbf{B}_{\phi}$ denote electric and magnetic part of $\phi$ field strength tensor $F_{\phi}^{\mu\nu}$, respectively.  Contributions from thermal vorticity $\omega$ and acceleration $\boldsymbol\varepsilon$ are also listed in Eq. (\ref{eq:rho_00}) for a comparison. Here all fields with primes are defined in the meson's rest frame. The coefficients $C_1$ and $C_2$ are constants that  depend only on the quark mass and the $\phi$ meson's mass  \cite{Sheng:2022wsy}
\begin{eqnarray}
C_{1} & = & \frac{8m_{s}^{4}+16m_{s}^{2}m_{\phi}^{2}+3m_{\phi}^{4}}{120m_{s}^{2}\left(m_{\phi}^{2}+2m_{s}^{2}\right)}\,,\nonumber \\
C_{2} & = & \frac{8m_{s}^{4}-14m_{s}^{2}m_{\phi}^{2}+3m_{\phi}^{4}}{120m_{s}^{2}\left(m_{\phi}^{2}+2m_{s}^{2}\right)}\,.\label{eq:C1/2}
\end{eqnarray}
In order to compare with the experimental data, the results in Eq. (\ref{eq:rho_00}) should be transformed back to the lab frame. Then the spin alignment in the lab frame are expressed in terms of fields in the lab frame, while the transformation depends on momenta of $\phi$ mesons. By taking average in the phase space, the mean value of spin alignment in the lab frame reads \cite{Sheng:2022ffb}
\begin{eqnarray}
\left\langle \rho_{00}^{\phi}\left(x,\mathbf{p}\right)\right\rangle _{x,\mathbf{p}} & \approx & \frac{1}{3}+\frac{1}{3}\underset{i=1,2,3}{\Sigma}\left\langle I_{B,i}(\mathbf{p})\right\rangle_p \frac{1}{m_{\phi}^{2}}\left[\left\langle \omega_{i}^{2}\right\rangle_x -\frac{4g_{\phi}^{2}}{m_{\phi}^{2}T_{\text{eff}}^{2}}\left\langle \left(\mathbf{B}_{i}^{\phi}\right)^{2}\right\rangle_x \right]\nonumber \\
 &  & +\frac{1}{3}\underset{i=1,2,3}{\Sigma}\left\langle I_{E,i}(\mathbf{p})\right\rangle_p \frac{1}{m_{\phi}^{2}}\left[\left\langle \varepsilon_{i}^{2}\right\rangle_x -\frac{4g_{\phi}^{2}}{m_{\phi}^{2}T_{\text{eff}}^{2}}\left\langle \left(\mathbf{E}_{i}^{\phi}\right)^{2}\right\rangle_x \right]\,,\label{eq:rho00_p}
\end{eqnarray}
where $\left\langle \cdots\right\rangle_x$,  $\left\langle \cdots\right\rangle_p$  denote average in space-time and in momentum space, respectively. The momentum-dependent functions $I_{E/B,i}(p)$ can be found in \cite{Sheng:2022ffb}. Equation (\ref{eq:rho00_p}) has a perfect factorization of $x$ and $\mathbf{p}$ dependence. Meanwhile, all field components appear in squares, indicating that the spin alignment measures the fluctuations of the fields instead of their mean values. 

\begin{figure}
\begin{center}
\includegraphics[width=0.47\textwidth]{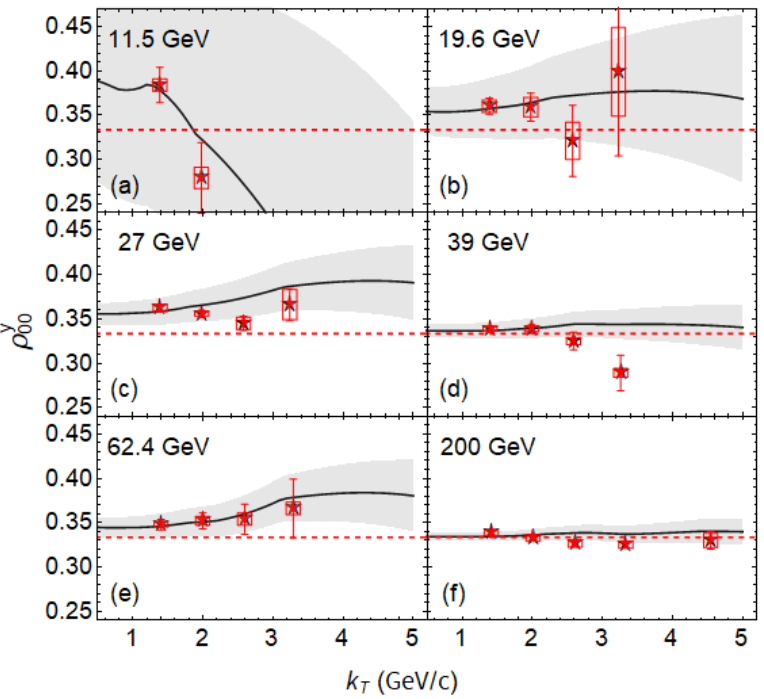} \includegraphics[width=0.49\textwidth]{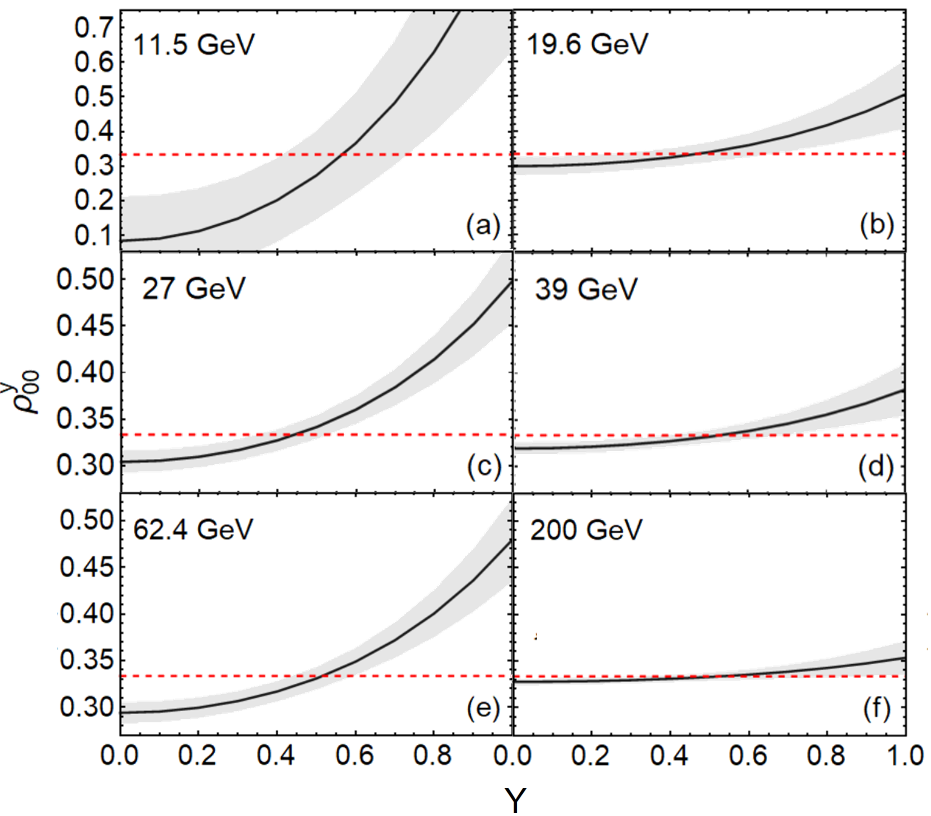}
\end{center}

\caption{Left panel: model predictions and uncertainties \cite{Sheng:2022wsy} (solid line with shaded areas) for the spin alignment $\rho_{00}$ of $\phi$ meson with respect to the event plane. The red stars with error bars are experiment results by the STAR collaboration \cite{STAR:2022fan}. Right panel: model predictions \cite{Sheng:2023urn} and uncertainties (solid line with shaded areas) for the spin alignment $\rho_{00}$ of $\phi$ meson as functions of rapidity. } \label{fig:rho00-kt}
\end{figure}

Due to the lack of theoretical inputs for the fluctuations, the authors in \cite{Sheng:2022wsy} choose to extract the transverse fluctuation $F_{T}^{2}=\left\langle \mathbf{E}_{x,y}^{2}\right\rangle =\left\langle \mathbf{B}_{x,y}^{2}\right\rangle$ and longitudinal fluctuation $F_{z}^{2}=\left\langle \mathbf{E}_{z}^{2}\right\rangle =\left\langle \mathbf{B}_{z}^{2}\right\rangle $ as functions of collision energies from experimental data for the $\phi$ meson's spin alignment in in-plain and out-of-plain directions \cite{STAR:2022fan}. With these parameters, the transverse momentum spectra of $\phi$ meson's spin alignment is calculated in \cite{Sheng:2022wsy}. The results are shown in the left panel of Fig. \ref{fig:rho00-kt}, where the experiment datas are plotted by red stars with error bars, while the model predictions are solid lines with shaded areas for uncertainties. The rapidity dependence of $\rho_{00}$ for $\phi$ mesons has also been predicted in \cite{Sheng:2023urn}, with the major results shown in the right panel of Fig. \ref{fig:rho00-kt}.

\section{Summary}

We briefly introduced the history of studies on polarizations in heavy ion collisions. The polarization for quarks is induced by nonlocal parton scatterings through the spin-orbit coupling, which can be analytically described by spin kinetic equations based on the Wigner function approach. At the hadronization stage, polarized quarks form hadrons and mesons. During this quark recombination process, the angular momentum is converted to the polarization of hadrons or vector mesons. For $\Lambda$ hyperons, the spin polarization is mainly carried by the constituent $s$ quark, whose global polarization is dominated by the vorticity field and the local polarization is dominated by the thermal shear tensor. On the other hand, the spin alignment of vector mesons may be mainly induced by fluctuating strong force field, which is generated by currents of pesudo-Goldstone bosons at the hadronization stage.

However, there are still some open questions waiting to be answered in the future. For the global or local polarization of hyperons, we are still lack of a comprehensive simulation based on spin kinetic theories or spin hydrodynamics which includes non-equilibrium effects. For vector mesons, we are wondering about possible connection between the spin alignment with the gluon condensates or glasma fields in the initial stage of heavy-ion collisions \citep{Kumar:2023ghs}.  The contribution from strong force field \cite{Sheng:2022wsy} and some other contributions from hydro quantities \cite{Li:2022vmb,Dong:2023cng,Wagner:2022gza}, to the spin alignment also need detailed and comprehensive quantative studies in the future. 

\bibliographystyle{JHEP}
\bibliography{ref}

\end{document}